\documentclass[12pt, letterpaper, onecolumn, draftcls]{IEEEtran}

\usepackage{indentfirst}
\usepackage{amsmath}
\usepackage{amssymb}
\usepackage{amsfonts}
\usepackage{dsfont}
\usepackage{cite}
\usepackage{times}
\usepackage[dvips]{graphicx}
\usepackage{comment}
\usepackage{epsf}
\usepackage{psfrag}
\usepackage{theorem}
\usepackage{geometry}
\usepackage{mathrsfs}
\usepackage{bbm}

\newcommand{\figw}{0.55\columnwidth}

\usepackage{acronym}
\begin{document}
\sloppy
\title{Relaying strategies for Uplink\\ in Wireless Cellular Networks}

\author{\authorblockN{{\bf Federico Librino}, {\bf Michele Zorzi}}\\
E--mail: \texttt{federico.librino@iit.cnr.it, zorzi@dei.unipd.it \vspace{-2mm}}
}

\date{}
\maketitle


\begin{abstract}
In this paper, we analyze the impact of relays on the uplink performance of FDMA cellular networks. We focus our analysis on Decode and Forward techniques, with the aim of measuring the improvements which can be achieved in terms of throughput and energy saving. We apply a stochastic geometry based approach to a scenario with inter-cell interference and reuse factor equal to 1.
The first goal of this work is to observe what is the impact of various relay features, such as transmission power, location and antenna pattern, when a half-duplex constraint is imposed.
The second goal is to determine how much relaying can be beneficial also for users who are not at the cell edge, and who can therefore use a direct link towards the base station. We show that if more refined decoding techniques, such as Successive Interference Cancellation and Superposition Coding, are properly used, considerable gains can be obtained for these mobiles as well.
\end{abstract}

\section{Introduction}
In recent years, the growing demand of advanced services and applications designed for the new generation of mobile phones has dramatically increased the traffic load which needs to be handled by cellular networks. Advanced applications are being currently run on mobiles, making it necessary for them to exchange a considerable amount of data to and from the Internet.
As a consequence, an efficient utilization of the available bandwidth is a critical need in future cellular networks. The development of HSDPA and HSUPA in 3G CDMA-based network paved the way for the LTE standard, which explicitly aimed at improving the data-oriented communications between mobiles and base stations.

The downlink is commonly regarded as the most demanding communication direction, due to the nature of most current applications: the choice of OFDMA in fact was meant to optimize the resource allocation in a highly loaded scenario. The organization of the communications through time-frequency resource elements makes it possible to better balance the resource sharing based on the different qualities of service required by the active UEs. The usage of MIMO techniques to exploit spatial diversity was also suggested with the same purpose.

Despite the large amount of research focused on downlink, the uplink also plays an important role, especially with the spread of applications such as file sharing or peer-to-peer. Due to the limited transmission power of mobiles, uplink transmission reliability often poses a limit to the coverage area of a single cell. In LTE, SC-FDMA is used for communications between the mobile terminals and the base stations, thus reducing the impact of interference within a single cell. However, inter-cell interference, especially if the frequency reuse factor approaches 1, is still present, and can severely hamper the uplink transmission reliability.

In order to cope with this issue, the latest release of the LTE standard also allows the deployment of wireless relays to help the cell-edge mobiles. The advantages of relaying have been widely investigated in the literature, and are obtained exploiting the broadcast nature of the wireless channel~\cite{mio22,mio7}. In fact, properly deployed nodes (which may also be, in principle, either regular terminals able to relay data packets or dedicated relays) can overhear surrounding transmissions, and therefore improve the link quality by adding spatial diversity. This in turn can be beneficial to counteract the effect of a highly dynamic interference level.

The basic relaying schemes, such as Amplify and Forward (A\&F) or Decode and Forward (D\&F), rely on the retransmission, performed by the relay, of the packets received from the source node~\cite{mio7,altroKramer,mio33,altroHost}. In the former case, the received signal (including interference and noise) is simply amplified before retransmission, while in the latter the packet is first decoded, and the obtained symbols are then re-encoded and transmitted. Clearly, D\&F requires additional complexity, but is also able to grant a higher benefit, since interference and noise amplification is avoided.

In cellular networks, fixed relays are usually not computationally limited, and this is why we focus on D\&F in this paper.
Furthermore, not only can they reduce the access link length, if properly located, but they may also take advantage of more effective antennas, due to their static position, and of a higher transmission power.
One of the common constraints which has to be considered, however, is the fact that relays are not able to receive and transmit simultaneously, at least on the same carrier, due to self-interference. This means that either time division or frequency division has to be adopted between the link from the mobile and the one towards the base station.
In the former case, unless adaptive modulation and/or coding is exploited, the advantage granted by the relay may be partially lost due to the reduced transmission time available; in the latter case, additional bandwidth is required, which may be a problem in congested scenarios.
Several schemes have been proposed recently, although most of them focus on downlink transmissions~\cite{relOFDM,oppOFDM,RelSel,Opt_powRelSel}. In~\cite{relOFDM}, practical problems about resource allocation are investigated, when OFDM is used; a distributed approach is presented to perform both intra-cell and inter-cell routing. A similar scenario is analyzed also in~\cite{oppOFDM}, where direct or relayed transmissions are properly scheduled with either spectrum reuse among sectors or a TDMA approach. A TDMA scheme, necessary to keep the communications from the BSs and from the relays orthogonal, is exploited in~\cite{RelSel} too; the problem of relay selection is addressed here in order to improve the overall cell throughput while maintaining the cell-edge users performance comparable to that of existing schemes.
The possibility of leveraging on mobiles as relays is considered in~\cite{Opt_powRelSel}, where a comparison between A\&F and D\&F is shown, and a joint optimization of power allocation, relay selection and relay strategy choice is developed.
An approach based on stochastic geometry has been used instead in~\cite{Baccelli} to analytically derive the coverage and the average rate in heterogeneous cellular networks, where however relays are not considered.

As to the Uplink, in~\cite{cell_ULrel} a mathematical derivation of the required mobile transmission power shows the benefit of uplink relaying in terms of energy saving. A scheme based on A\&F and on Compress and Forward (C\&F) is instead described in~\cite{cell_UL_fulldup}, which however is designed mainly for cooperating full-duplex relays and non fading channels.
The possibility of exploiting other mobiles as relays in CDMA based networks has been detailed in~\cite{cell_UL_CDMA}, whereas D\&F through both fixed and mobile relays is investigated from an information theoretic perspective in~\cite{cell_UL_bord}, with focus on cell-edge users only.
Channel correlation between cooperative links, when mobiles are used as relays, is considered in~\cite{cell_UL_corr}, together with relay selection, although interference is neglected. In~\cite{Pow_RelSel}, relays help cell-edge users; power allocation and relay selection are performed jointly through a distributed scheme, also when Channel State Information is limited or imperfect. However, no inter-cell interference is considered.

In general, two aspects are rarely taken into account when relaying is studied for the uplink.
First, although it is intuitive that relaying can be beneficial for mobiles located too far from the base station, it is not clear how much this can be true for the remaining users and, therefore, whether it is really worth or not to deploy relay stations. Secondly, the analyzed scenario is often restricted to a single cell, despite the fact that cell-edge users, which are the main target when using relaying, may often be severely damaged by interference coming from surrounding cells.

In this work, we focus on the usage of TDD relays for the uplink of an FDMA-based cellular network. We want to investigate the improvement, in terms of outage probability, which can be offered through relaying when advanced receiving techniques are available. In addition, our aim is also to determine whether and how much relays can help users which are not necessarily outside the coverage range, and to verify the effectiveness of relaying when inter-cell interference is taken into account.

The main idea behind our study is that both the link from mobile to relay (\textit{access link}) and the one from relay to base station (\textit{backhaul link}) are usually strong enough to make it possible to simultaneously use also the direct link, from mobile to base station, if available. In doing this, we want to avoid the need for extra bandwidth, while compensating the half-duplex constraint of the relay.

In order to simultaneously use the direct and backhaul links Interference Cancellation is a promising approach. First described in~\cite{BroadChannel}, Successive Interference Cancellation (SIC) has been widely investigated and applied in different scenarios, such as ad hoc networks~\cite{Info_SIC}, WCDMA cellular networks~\cite{algo_SIC}, and even for reducing collisions in ALOHA-based protocols for satellite communications~\cite{SIC_CRDSA}.
Intra-cell interference cancellation can grant up to a 50\% capacity increase for the downlink, as shown in~\cite{algo_SIC}, while the importance of accurate cancellation of at least the strongest interferer has been clarified in~\cite{Info_SIC}. Exhaustive descriptions of SIC schemes for MIMO systems, which consider also inter-cell interference, are collected in~\cite{DL_MIMO_SIC} and can be also found in~\cite{MIMO_SIC_survey} and in the references therein.
Recently, an overview of possible IC mechanisms for the Uplink, as well as some comments on future directions for this technology in 4G networks, have been presented in~\cite{SIC_cell}. Directional antennas, as well as various channel models (Rayleigh, Nakagami, Rice) have been also analyzed in~\cite{Cell_SICperf}.
In general, Successive Interference Cancellation is particularly suited when the two incoming signals to be decoded show significantly different power levels. This usually happens when relaying is exploited for cell edge users, but the higher transmission power of the relay makes it true also for mobiles closer to the base station.

Superposition Coding~\cite{BroadChannel}, conversely, is an effective way to deliver multiple data streams to different receivers, by properly sharing the transmit resources between them. It is therefore suited to communicate over the direct and the access link at the same time, since the information on the two channels is usually available at the mobile terminal.

Interference Cancellation and Superposition Coding can be seen as complementary techniques, whose dual behavior makes it possible also to combine them together, as proposed in~\cite{SC_SIC_exp,SC_HARQ,Dave_SC_cog,SC_relay}. In particular, the \textit{SC-relaying} proposed in~\cite{SC_relay} is close to the idea proposed in this work, although the aim of the authors there is to generalize the problem of adaptive modulation and coding, and they do not consider interference from surrounding nodes.

To sum up, the main contributions of this paper are:
\begin{itemize}
 \item we analyze the effectiveness of relaying schemes for uplink in FDMA cellular networks, by taking into account also the inter-cell interference by means of a Stochastic Geometry approach;
 \item we observe how much relaying is beneficial when mobiles are not necessarily at the cell-edge. To this aim, Successive Interference Cancellation and Superposition Coding are combined to derive more flexible relaying schemes;
 \item we analyze the impact on the overall performance of some key relay features, such as transmission power, location, and directional antennas, as well as of some relay constraints, such as half-duplex limitation;
\end{itemize}


The paper is organized as follows. In Section~\ref{sys_mod} we describe the considered system model, and the mathematical tools used to model the SIC and SC procedures. Section~\ref{out_prob} derives the outage probability expressions for all the considered transmission schemes, from which throughput and energy efficiency are computed, when SIC only is employed. The advantages granted by SC are instead discussed in Section~\ref{adv_SC}, whereas the results are presented in Section~\ref{results}. Section~\ref{concl} concludes the paper.

\section{System Model}
\label{sys_mod}
In this paper, we study the performance of cellular networks when relays are adopted to improve the performance in terms of throughput and energy consumption.
The SC-FDMA scheme adopted in 4G cellular networks is considered, with frequency reuse factor equal to 1, resulting in one user allocated to each BS per carrier.
We focus on a single carrier, since the extension to multi-carrier just requires to repeat the same derivation in each of the other orthogonal channels and to combine the results.

We investigate the achievable throughput and energy consumption in a given cell. However, since we want to evaluate also the impact of inter-cell interference, we need to model the locations of the mobiles allocated to the same carrier in the other cells.
Different approaches can be adopted in doing this, and two of them could be considered in particular.
With the first approach, we distribute the BSs according to a Poisson Point Process (PPP) $\mathcal{P}=\{b_1,b_2,\ldots\}\in\mathbb{R}^2$ of intensity $\lambda$, and focus on a specific one. Successively, we identify the Voronoi cells corresponding to the deployed BS, and distribute one mobile in each cell. As a result, each mobile terminal is allocated to its closest BS.

A second approach is to focus on a given BS, and to deploy instead the UEs as a Poisson Point Process (PPP) $\mathcal{P}=\{b_1,b_2,\ldots\}\in\mathbb{R}^2$ of intensity $\lambda$. The closest UE to the BS is the one allocated within the cell of interest, whereas the remaining ones are the interferer UEs of the other cells.


The two models are not equivalent. Two main differences can be observed. The former is the distribution of the distance between the BS and its allocated mobile, the latter is the distribution of the distance between the BS and the closest interferer.
Simulations show that both distributions are shifted towards higher values with the second model, which can lead to slightly worse performance, especially when the node density is very low.
Nevertheless, comparisons by simulation showed that, with the parameters setup of all the investigated scenarios, the performance obtained with the two models is almost the same.
For this reason, and since the second model allows for a more tractable analysis, we adopt it in the following.

%

The channel is modeled by taking into account both the path loss contribution $p_{\ell}(d) = A d^{-\alpha}$, which depends on the distance $d$ between the mobile and the BS, and the fading contribution $h(t)$, which is time varying,
and is modeled as an exponential random variable of unit power (Rayleigh fading). $A$ and $\alpha$ are fixed positive numbers. Temporal correlation of $h(t)$ could also be introduced in our model, as well as spatially-correlated shadowing effects, but with no substantial changes in the derivations. Therefore, for simplicity, we do not consider them in this work.

As regards the relays, we assume that each cell is also equipped with a fixed number $k_r$ of relays, which are deployed at a fixed distance $d_{rb}$ from the BS, with equal angular separation. The relays are different from the common UEs, and operate to help users by means of a Decode and Forward (D\&F) scheme. Not only is their transmit power higher, but, given their fixed position, they can also make use of a very directive antenna towards the corresponding BS.
The relay transmit power $\bar{P}_r$, multiplied by the antenna gain $\eta$, corresponds to the \textit{effective transmit power} $P_r$, which is usually higher than $P_t$. In addition, we consider that the directional antenna used on the backhaul link (i.e., from relay to BS) is enough to prevent the relay from interfering with other BSs, or to make such interference effect negligible.
The half-duplex constraint prevents relays from receiving and transmitting simultaneously.

In order to consider a more general case, we add the option, for both the relays and the BSs, of using directional antennas also to receive signals on the access link (from UE to relay) and on the direct link (from UE to BS), respectively. We do not make any assumption on how directional antennas are formed, if electronically obtained or mechanically designed.
Any kind of beam pattern can be considered. We choose here to test a simple pattern with a single main lobe. The cosine model~\cite{cosine} is commonly adopted for numerical simulations: if $-\pi\leq\theta\leq\pi$ is the angle with respect to the antenna direction, we define:
\begin{equation}
 f(\theta) = \left(\frac{1+\cos(\theta)}{2}\right)^k
\label{trigopattern}
\end{equation}
as the antenna pattern, where $k\geq 0$ is a parameter to tune the antenna directivity. Notice that setting $k=0$ makes the antenna omnidirectional.

We assume that time is divided into slots of fixed duration, and that slot synchronization is available at all nodes, as commonly verified in practical cellular networks. A data packet is sent by each UE in every time slot. We define a threshold value $\vartheta$ for the Signal to Interference plus Noise Ratio (SINR), such that a packet is correctly decoded whenever the SINR of its transmission is higher than $\vartheta$, and found in error otherwise. With this model, the decoding probability in a given time slot for a transmission performed over the link between $i$ and $j$ is:
\begin{equation}
 P_s = \mathbb{P}\left[\frac{P_ip_{\ell}(d_{ij})h_{ij}(t)f(\theta_{ji})}{N_0+I} \geq \vartheta\right]
\end{equation}
where $P_i$ is the transmit power (equal to $P_t$ or to $P_r$), $d_{ij}$ is the distance between the two terminals, and $\theta_{ji}$ is the angle between the direction of the antenna of the receiving node $j$ and the direction of the line joining $j$ and $i$. No factors for the trasmitting antenna are required. In fact, if $i$ is a UE, its antenna is omnidirectional; if instead it is a relay, the value of the antenna gain is already taken into account in the effective transmission power. In the denominator, $N_0$ is the noise power, whereas $I$ is the overall interference level from the other cells in the network.

\subsection{Interference Cancellation scheme}
\label{sec:intcanc}
When more refined decoders are employed at the Base Stations, it is possible to consider a system which leverages Successive Interference Cancellation (SIC). In this work, we assume that up to two incoming transmissions can be decoded successively, and that cancellation is error free. If we suppose that signal 1 is decoded first, the probability of retrieving both data packets becomes:
\begin{equation}
 P_{ic}^{(12)} = \mathbb{P}\left[\frac{\rho_1}{N_0+\rho_2+I_1}\geq\vartheta, \frac{\rho_2}{N_0+I_2}\geq\vartheta\right]
\end{equation}
where $\rho_1$ and $\rho_2$ are the received powers of the two incoming signals. An analogous expression is applied if we assume that signal 2 is instead decoded first. The two expressions are used to derive the unconditioned probability of decoding both packets.
Usually, the interference levels $I_1$ and $I_2$ are the same, but this is not necessarily true, as may be the case for an electronically formed directional antenna (for instance, through MIMO processing), which may process the received signal with different weighs (corresponding to different antenna directions), resulting in different interference levels for different intended signals.

Interference cancellation can be applied also to the signal coming from a single source, if multiple streams are sent together. This may be achieved, for instance, through signal multiplexing in MIMO transmitters. If a single antenna is available, as is the case for mobile terminals, other techniques can be applied, like \textit{Superposition coding}.

\subsection{Superposition Coding}
If SIC is exploited at the receiver, it is possible to take advantage also of a more refined transmission scheme, implementing the Superposition Coding principle (SC). The main idea behind this technique is to divide the resources (time, power) of the transmitter to handle the transmission of a higher number of data streams.
At the receiver, the two streams are successively decoded as described in the previous section: the one which was assigned the highest resource fraction is decoded first, with the other one as interference. The only difference is that both streams are sent over the same channel. In case two packets are superimposed, with a fraction $1/2\leq\beta<1$ of the power reserved to packet 1 and a fraction $1-\beta$ to packet 2, the probability of decoding both of them is:
\begin{eqnarray}
 P_{sc}^{(12)} & = & \mathbb{P}\left[\frac{\beta \rho}{(1-\beta)\rho+I+N_0}\geq\vartheta, \frac{(1-\beta)\rho}{I+N_0}\geq\vartheta\right]
\end{eqnarray}
where $\rho$ is the overall power received at the destination node from the transmitter, and perfect cancellation is assumed, as above. It is possible that only the first packet can be recovered; however, its succesful decoding is necessary for the latter to be also correctly received.
Clearly, a careful tuning of $\beta$ is a key issue in determining the effectiveness of the scheme.

\subsection{Relaying schemes}
We describe here the four communication schemes which will be evaluated in the following sections.
Given the half-duplex constraint at the relay, it follows that any relayed packet requires two time slots to be delivered to the BS.
Therefore, for a fair comparison, each scheme encompasses two subsequent time slots.
Notice that, for the schemes based on relaying, despite the better channel conditions of the backhaul link, and the reduced lengths of the links, the resulting gain must be large enough to compensate the double amount of time required to deliver the packet.

In order to face this problem, we assume that a Successive Interference Cancellation scheme is available for decoding at the Base Station, as described in Section \ref{sec:intcanc}.

The \textit{Basic} scheme does not exploit the presence of relays. Therefore, it simply consists of a transmission of a packet from the UE to the BS in each of the two time slots.

\begin{figure}
    \centering
    \includegraphics[width=\figw]{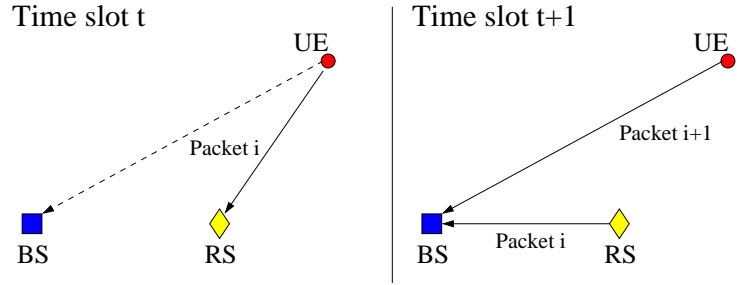}
     \caption{\small Scheme of how the relay is used in the proposed relaying schemes.
     }
     \vspace{-1cm}
  \label{fig:ProtoScheme}
\end{figure}

The \textit{Baseline Relaying} scheme, on the contrary, takes advantage of the presence of the relay station, together with the SIC scheme at the BS. With this approach, the two time slots are utilized as follows:
\begin{itemize}
 \item in time slot 1, only the UE transmits towards the relay, which tries to decode the packet;
 \item in time slot 2, the relay transmits the packet received in the previous time slot, if decoding was successful; at the same time, the UE transmits a new packet towards the BS, which performs SIC to retrieve both packets. If the relay did not succeed in decoding the UE transmission, then only the UE transmits to the BS in this second time slot.
\end{itemize}
In this scheme, depicted in Figure \ref{fig:ProtoScheme}, we assume therefore that the relay transmits whenever it receives a packet in the first time slot, and that the BS does not use the direct link. This is what happens in the widely studied scenario where the link between UE and BS is very bad.
We start with this scheme because it is easier to analyze formally, and the results can be applied, with minor modifications, to the two following ones as well.

The proposed baseline relaying scheme is not always very effective if the direct link can be used. In fact, the help of the relay is fundamental for UEs located far from the BS, but it may become useless or even detrimental if the UE is close to the BS. Since we are interested in assessing the improvement granted by relays also when the direct link is available, we can easily add two refinements.

If the BS also listens to the direct link, it can exploit the spatial and temporal diversity by storing, in case of successful decoding, the packet transmitted by the UE in the first time slot, which corresponds to selecting the best path between direct and relayed. We call this scheme \textit{Selection Relaying}.

If feedback can be sent on the downlink channel towards the relay, the BS can prevent the relay from transmitting in case it has already received the packet from the UE, thus avoiding its interference when the UE performs its second transmission during the second time slot. We obtain the \textit{Feedback relaying} scheme.
Both these improvements are likely to affect the UE if it is located in the center of the cell, since it actually does not always need the help of the relay.

Superposition coding may also be beneficial. If it is available, the four schemes are slightly modified: in the \textit{Basic} scheme, the UE always transmits two superimposed packets; in the relaying schemes, superposition coding is used only in the first UE transmission, with no other modifications.

In the following section, we first analyze the four schemes to assess their performance in terms of throughput and energy consumption. As a second step, we observe the same performance metrics when Superposition Coding is exploited.

\section{Outage probability analysis}
\label{out_prob}
In this section, we focus on the outage probability of the four schemes, which will then be used to determine the effectiveness of relayed transmissions in terms of throughput and energy consumption, following an approach similar to \cite{Haenggi, relOFDM}.
In all the schemes, the interference is due to the transmissions of the UEs of the other cells. Such interferers can be assumed to be distributed as a PPP, as explained before. Due to Campbell's Theorem \cite{Mecke}, we define two quantities which will be useful in the following.

The first one is related to the Laplace transform of the interference level, expressed as:
\begin{eqnarray}
 \mathscr{L}(s,x) & = & \mathbb{E}\left[e^{-sI}\right] \nonumber\\
 & = & \exp\left(-\lambda\int_0^{2\pi}\int_x^{+\infty}\left(1-\frac{1}{1+sP_tAr^{-\alpha}}\right)r\,drd\theta\right) \nonumber\\
 & = & \exp\left(-2\pi\lambda\int_x^{+\infty}\frac{sP_tAr}{r^{\alpha}+sP_tA}\,dr\right) \nonumber \\
 & = & \exp\left(-2\pi\lambda sP_tA\frac{x^{2-\alpha}}{\alpha-2}\,\,{}_2F_1\left(1,1-\frac{2}{\alpha};2-\frac{2}{\alpha};-s\frac{P_tA}{x^{\alpha}}\right)\right)
\end{eqnarray}
where ${}_2F_1(a,b;c;d)$ is the hypergeometric function\cite{inte}.

Since the Poisson Process is homogeneous, the Laplace transform does not depend on the location where the interference level is measured, but depends on the distance $x$ between the transmitter and the receiver since, by hypothesis, the transmitter is the closest UE to the BS.

The second quantity is the joint Laplace transform of the interference levels $I_1$ and $I_2$, measured in two locations whose distance is $d$. By choosing a reference system such that the two considered points are located at $(d,0)$ and at $(0,0)$, respectively, we compute:
\begin{eqnarray}
 \mathscr{L}_d(s,t,x) & = & \mathbb{E}\left[e^{-sI_1}e^{-tI_2}\right] \nonumber\\
 &\hspace{-4.5cm} = &\hspace{-2.5cm} \exp\left(-\lambda\int_0^{2\pi}\int_x^{+\infty}\left(1-\frac{1}{\left(1+sP_tA\left(r^2+d^2-2rd\cos(\theta)\right)^{-\alpha/2}\right)\left(1+tP_tAr^{-\alpha}\right)}\right)r\,drd\theta\right)
 \label{defLd}
\end{eqnarray}
which in general cannot be rewritten in closed form, but can be easily computed numerically. We point out that in both definitions we assumed that all the interferers transmit with the same power $P_t$, according to our assumptions above.

\subsection{Throughput computation}
For the throughput derivation, we consider a BS located in $(0,0)$, and a relay located in $(d_{rb}, 0)$. If $k_r$ relays per BS are deployed, the UE distance $d_{ub}$ from the BS is Poisson distributed, whereas the angle $\theta_u$ is uniformly distributed between $-\pi/k_r$ and $\pi/k_r$. The distance $d_{ur}$ between the UE and the relay can be obtained geometrically.

For the sake of clarity, we list the following definitions:
\begin{itemize}
 \item we call $\gamma_{ij}$ the average SNR between terminals $i$ and $j$, for $i,j\in\{u,r,b\}$. Correspondingly, we have $\gamma_{ub} = AP_t/(N_0d_{ub}^{\alpha})$, $\gamma_{rb} = AP_r/(N_0d_{rb}^{\alpha})$ and $\gamma_{ur} = AP_t/(N_0d_{ur}^{\alpha})$. Notice that interference is not considered here;
 \item we call $\chi_{ij}$ the indicator random variable whose value is 1 if a packet is correctly transmitted from node $i$ to node $j$, with $i,j\in\{u,r,b\}$. Notice that while $\chi_{ur}$ can be easily derived, since IC is not required at the relay, the expression of $\chi_{rb}$ must take into account that a concurrent transmission from the UE to the BS is present. Finally, as to the channel between the UE and the BS, we consider that $\chi_{ub}$ is used when there is no simultaneous transmission from the relay;
 \item we call $\chi_{ub}^i$ the indicator random variable indicating that a packet from the UE is correctly decoded at the BS when a concurrent transmission from the relay is also ongoing.
\end{itemize}
It is worth noting that the random variables $\chi_{ij}$ are not independent, since the interference is spatially correlated. All the expectations of these random variables are taken over both the channel coefficient and the interference level.
Since the relaying scheme needs two time slots to deliver the relayed packet, we measure the throughput as the average number of packets received at the BS during two subsequent time slots.

\subsubsection{Basic scheme}
Since omnidirectional antennas are considered now, for the throughput of the basic scheme we only need to know the distance $d_{ub}$. 
During two subsequent time slots, two direct transmissions are performed.
The probability of correct reception of a direct transmission is:
\begin{eqnarray}
 \mathbb{E}_{h,I}\left[\chi_{ub}\right] & = & \mathbb{E}_{h,I}\left[\mathbbm{1}\left(SINR \geq \vartheta\right)\right]\nonumber\\
 & = & \mathbb{E}_I\left[P\left[\frac{P_tAd_{ub}^{-\alpha}}{N_0+I}h \geq \vartheta\right]\right] \nonumber\\
 & = & \mathbb{E}_I\left[P\left[h\geq \frac{\vartheta}{\gamma_{ub}} + \frac{\vartheta}{P_tA}d_{ub}^{\alpha}I \right]\right] \nonumber\\
 & = & e^{-\frac{\vartheta}{\gamma_{ub}}}\mathscr{L}\left(\frac{\vartheta}{P_tA}d_{ub}^{\alpha},d_{ub}\right)
 \label{Throdir}
\end{eqnarray}
where we used the fact that $h$ is exponentially distributed with parameter $1/\gamma_{ub}$.

We notice that, although the computation of the probability that both transmissions are successful requires to take into account the interference correlation, the average throughput can be computed by looking only at a single time slot. The corresponding throughput is in fact $T_{dir}(d_{ub}) = 2\,\mathbb{E}_{h,I}\left[\chi_{ub}\right]$.

\subsubsection{Baseline Relaying scheme}
We now move to the computation of $T_R$, that is, the average throughput achieved by exploiting relaying and Interference Cancellation at the BS.

As a first step, we recall the decoding probability expression when SIC is used at the BS. We call $\Gamma_{rb}$ and $\Gamma_{ub}$ the exponential random variables representing the received SNRs of the transmissions from the relay to the BS and from the UE to the BS, whose parameters are $1/\gamma_{rb}$ and $1/\gamma_{ub}$, respectively.

We first assume that the BS is able to decode the signals starting from the strongest one. The probability that both signals are successfully decoded hence reads as:
\begin{equation}
 P\left[\frac{\Gamma_{rb}}{\Gamma_{ub}+\hat{I}+1}\geq\vartheta, \frac{\Gamma_{ub}}{\hat{I}+1}\geq\vartheta, \Gamma_{rb}\geq \Gamma_{ub}\right] + P\left[\frac{\Gamma_{ub}}{\Gamma_{rb}+\hat{I}+1}\geq\vartheta, \frac{\Gamma_{rb}}{\hat{I}+1}\geq\vartheta, \Gamma_{rb}< \Gamma_{ub}\right]
 \label{decoSIC}
\end{equation}
where $\hat{I}=I/N_0$ is the value of the interference normalized to the noise power. The expressions of the two terms in (\ref{decoSIC}) are symmetric, so we just derive the first one. Since we focus on the case $\vartheta\geq 1$, we can ignore the constraint $\Gamma_{rb}\geq \Gamma_{ub}$. From the distribution of the random variables $\Gamma_{rb}$ and $\Gamma_{ub}$ we can easily derive, for a given value of $\hat{I}$:
\begin{equation}
 P\left[\frac{\Gamma_{rb}}{\Gamma_{ub}+\hat{I}+1}\geq\vartheta, \frac{\Gamma_{ub}}{\hat{I}+1}\geq\vartheta\right] = e^{-\vartheta(\hat{I}+1)\left(\frac{1}{\gamma_{rb}} + \frac{1}{\gamma_{ub}}\right)}\frac{e^{-\vartheta^2(\hat{I}+1)/\gamma_{rb}}}{1+\frac{\gamma_{ub}}{\gamma_{rb}}\vartheta}
 \label{prob_dec_IC}
\end{equation}
and a symmetric expression is obtained for the second term. The interference term $\hat{I}$ is finally to be averaged. In order to better understand the expression in (\ref{prob_dec_IC}), we notice that:
\begin{itemize}
 \item the first exponential is due to the fact that the SINR of both signals must be greater than $\vartheta$;
 \item the second exponential derives from the fact that the received signal from the relay must be greater than the one from the UE, which in turn must be greater than $\vartheta(I+N_0)$, as stated above. This actually implies that the received signal from the relay must be greater than $\vartheta(\vartheta+1)(I+N_0)$. This term disappears if we do not want the signal from the UE to be also decoded;
 \item the denominator comes from the fact that when the signal from the relay is decoded, the interference from the UE is still to be taken into account. This term remains also if we do not want the signal from the UE to be above threshold.
\end{itemize}

For the throughput computation, we observe that the average throughput is given by the sum of two components, namely the average throughput over the direct link and the one over the relayed path, having in mind that each one may be affected by the interference of the other one at the BS. For a UE located at the position whose polar coordinates are $d_{ub}$ and $\theta_u$, we have:
\begin{eqnarray}
 T_R(d_{ub},\theta_u) & = & \frac{1}{2}\,\mathbb{E}[T_{dir}(d_{ub})] + \mathbb{E}[T_{rel}(d_{ub},\theta_u)] \nonumber\\
 & = & \mathbb{E}\left[\chi_{ur}\chi_{rb}\right] + \mathbb{E}\left[\chi_{ub}\right] - \mathbb{E}\left[\chi_{ur}\chi_{ub}\right] + \mathbb{E}\left[\chi_{ur}\chi_{ub}^i\right]
\label{thro_rel_nofeeed}
\end{eqnarray}

Equation (\ref{thro_rel_nofeeed}) can be explained as follows. The term $\chi_{ur}\chi_{rb}$ expresses the average throughput on the relayed path, which requires both access and backhaul link to be traversed. The second term is the throughput on the direct link, which is used by the second packet sent by the UE. However, this throughput does not consider the interference from the relay, which occurs when the relay successfully receives the first packet.
If this is the case, the throughput on the direct link is instead determined by $\chi_{ub}^i$. This explains why we need to subtract the term $\mathbb{E}\left[\chi_{ur}\chi_{ub}\right]$ and add $\mathbb{E}\left[\chi_{ur}\chi_{ub}^i\right]$ to retrieve the correct expression.
Notice that when the link from the UE to the relay is very good, such that $\mathbb{E}\left[\chi_{ur}\right]\rightarrow 1$, the throughput reduces to $\mathbb{E}\left[\chi_{rb}\right] + \mathbb{E}\left[\chi_{ub}^i\right]$, which is correct, since there is always interference on the direct link. On the contrary, if $\mathbb{E}\left[\chi_{ur}\right]\rightarrow 0$ (very bad access link), the throughput is equal to $\mathbb{E}\left[\chi_{ub}\right]$, which is exactly half the value obtained in the basic scheme, without the help of the relay.

All the expectations are taken over both the interference and the channel coefficients. As an example, we report the expression of $\mathbb{E}\left[\chi_{ur}\chi_{rb}\right] = P[\chi_{ur}=1,\chi_{rb}=1]$. The computation over the access link is easy, since there is no SIC at the relay. On the contrary, the decoding on the backhaul link is influenced by the presence of a symultaneous transmission from the UE to the BS.
Since decoding at the relay is performed after sorting the SINRs, two cases can occur. If the signal from the relay is stronger and decoded first, we can focus on its decoding only; if instead the signal from the UE is stronger, it is necessary to decode it too, since $\vartheta\geq1$, in order to cancel its interference. In this case, therefore, we need that both packets are correctly received.

Therefore, we have:
\begin{eqnarray}
 \mathbb{E}\left[\chi_{ur}\chi_{rb}\right] & = &  \mathbb{E}_{I_B,I_R}\left[P\left[\frac{\Gamma_{ur}}{\hat{I}_R+1}\geq\vartheta,\frac{\Gamma_{rb}}{\Gamma_{ub}+\hat{I}_B+1}\geq\vartheta\right]+\right.\nonumber\\
& & + \left.P\left[\frac{\Gamma_{ur}}{\hat{I}_R+1}\geq\vartheta,\frac{\Gamma_{ub}}{\Gamma_{rb}+\hat{I}_B+1}\geq\vartheta, \frac{\Gamma_{rb}}{\hat{I}_B+1}\geq\vartheta\right]\right] \nonumber\\
 & = & e^{-\vartheta\left(\frac{1}{\gamma_{ur}}+\frac{1}{\gamma_{rb}}\right)}\frac{1}{1+\frac{\gamma_{ub}}{\gamma_{rb}}\vartheta}\,\mathbb{E}_{I_B,I_R}\left[e^{-\frac{\vartheta}{\gamma_{ur}}\hat{I}_R}e^{-\frac{\vartheta}{\gamma_{ub}}\hat{I}_B}\right] + \nonumber\\
 & & + e^{-\vartheta\left(\frac{1}{\gamma_{ur}}+\frac{1}{\gamma_{rb}}+\frac{1}{\gamma_{ub}}\right)}\frac{e^{-\frac{\vartheta^2}{\gamma_{ub}}}}{1+\frac{\gamma_{rb}}{\gamma_{ub}}\vartheta}\,\mathbb{E}_{I_B,I_R}\left[e^{-\frac{\vartheta}{\gamma_{ur}}\hat{I}_R}e^{-\frac{\vartheta(\vartheta+1)}{\gamma_{ub}}\hat{I}_B}e^{-\frac{\vartheta}{\gamma_{rb}}\hat{I}_B}\right] \nonumber \\
 & \hspace{-4cm}= &\hspace{-2.3cm} \frac{e^{-\vartheta\left(\frac{1}{\gamma_{ur}}+\frac{1}{\gamma_{rb}}\right)}}{1+\frac{\gamma_{ub}}{\gamma_{rb}}\vartheta}\mathscr{L}\left(\frac{\vartheta}{N_0\gamma_{ur}},\frac{\vartheta}{N_0\gamma_{rb}}\right) + \frac{e^{-\vartheta\left(\frac{1}{\gamma_{ur}}+\frac{1}{\gamma_{rb}}+\frac{(\vartheta+1)}{\gamma_{ub}}\right)}}{1+\frac{\gamma_{rb}}{\gamma_{ub}}\vartheta}\mathscr{L}_{d_{rb}}\left(\frac{\vartheta}{N_0\gamma_{ur}},\frac{\vartheta}{N_0}\left(\frac{\vartheta+1}{\gamma_{ub}}+\frac{1}{\gamma_{rb}}\right)\!,d_{ub}\right)\nonumber\\
 \label{E_urrb}
\end{eqnarray}

The expression for $\mathbb{E}\left[\chi_{ur}\chi_{ub}^i\right]$ is exactly the same, but the values $\gamma_{rb}$ and $\gamma_{ub}$ are switched, since we want the packet from the UE to be decoded. Finally, the derivation of $\mathbb{E}\left[\chi_{ur}\chi_{ub}\right]$ is even simpler, since SIC is not involved, and is equal to:
\begin{eqnarray}
 \mathbb{E}\left[\chi_{ur}\chi_{ub}\right] & = & e^{-\vartheta\left(\frac{1}{\gamma_{ur}}+\frac{1}{\gamma_{ub}}\right)}\mathscr{L}_{d_{rb}}\left(\frac{\vartheta}{N_0\gamma_{ur}},\frac{\vartheta}{N_0\gamma_{ub}},d_{ub}\right)
\end{eqnarray}

\subsubsection{Selection Relaying and Feedback Relaying schemes}
The average throughput values $T_R^s(d_{ub},\theta_u)$ and $T_R^{f}(d_{ub},\theta_u)$ of a UE with polar coordinates $(d_{ub},\theta_u)$ can be derived as follows.

Since we have to take into account two transmissions over the direct link (in slot 1 and in slot 2), we rename the indicator variables $\chi_{ub}$ and $\chi_{ub}^i$ as $\chi_{ub2}$ and $\chi_{ub2}^i$, and introduce $\chi_{ub1}$, whose definition follows from that of $\chi_{ub}$, defined above. Notice that $\chi_{ub1}$ and $\chi_{ub2}$ have the same distribution. For the \textit{Selection Relaying} case, we need to add the average throughput over the direct link in the first time slot. However, since the same packet may be received also through the relayed path, we obtain:
\begin{equation}
 T_R^s(d_{ub},\theta_u) = T_R(d_{ub},\theta_u) + \mathbb{E}\left[\chi_{ub1}\right] - \mathbb{E}\left[\chi_{ub1}\chi_{ur}\chi_{rb}\right]
\end{equation}
which is higher than $T_R(d_{ub},\theta_u)$, since $\mathbb{E}\left[\chi_{ub1}\right] > \mathbb{E}\left[\chi_{ub1}\chi_{ur}\chi_{rb}\right]$.

Finally, in order to derive $T_R^f(d_{ub},\theta_u)$, we need to point out that it is possible to have a transmission over the direct link in the second time slot without interference from the relay even when the relay decoded the first packet, provided that the same packet has been decoded also at the BS during the first time slot. Therefore, we obtain:
\begin{equation}
 T_R^{f}(d_{ub},\theta_u) = T_R^s(d_{ub},\theta_u) - \mathbb{E}\left[\chi_{ub1}\chi_{ur}\chi_{ub2}^i\right] + \mathbb{E}\left[\chi_{ub1}\chi_{ur}\chi_{ub2}\right]
\end{equation}

The expressions of $\mathbb{E}\left[\chi_{ub1}\chi_{ur}\chi_{rb}\right]$, $\mathbb{E}\left[\chi_{ub1}\chi_{ur}\chi_{ub2}^i\right]$ and $\mathbb{E}\left[\chi_{ub1}\chi_{ur}\chi_{ub2}\right]$ can be obtained in a way similar to (\ref{E_urrb}), and are reported in the Appendix.

Having determined the average throughput $T(r,\theta)$ for a UE located at a given position $(r,\theta)$, where $T$ is equal to $T_{dir}$, $T_R$, $T_R^s$ or $T_R^f$, depending on the scheme, we finally compute the achievable throughput per cell per carrier by averaging over the user location distribution, as defined above:
\begin{equation}
 \bar{T} = \int_{-\frac{\pi}{k_r}}^{\frac{\pi}{k_r}}\int_{0}^{+\infty}T(r,\theta)2\pi\lambda re^{-\lambda\pi r^2}\frac{k_r}{2\pi}\,drd\theta = k_r\lambda\int_{-\frac{\pi}{k_r}}^{\frac{\pi}{k_r}}\int_{0}^{+\infty}T(r,\theta) re^{-\lambda\pi r^2}\,drd\theta
 \label{avethro}
\end{equation}

We can also numerically compute the throughput distribution. In fact, its cumulative distribution function $F_T(t)$ can be expressed as:
\begin{equation}
 F_T(t) = k_r\lambda\int_{-\frac{\pi}{k_r}}^{\frac{\pi}{k_r}}\int_{0}^{+\infty}\mathbbm{1}\left[T(r,\theta)\leq t\right] re^{-\lambda\pi r^2}\,drd\theta
 \label{disthro}
\end{equation}

\subsection{Energy consumption computation}
When relaying is used, a higher throughput can be achieved for cell edge users; however, this comes at the cost of additional energy spent at the relay. We investigate whether this additional energy cost is worth or not. We want to stress that the fixed position of the relay, which may be a limit, since it can help only a subset of the UEs, can be also seen as a strength, since a highly directional antenna can be used for the backhaul link. We represented this gain within the term $P_r$, defined as the effective transmit power of the relay. This term is in turn equal to $\eta \bar{P}_r$, where $\bar{P}_r$ is the actual power spent by the relay, and $\eta$ is the transmit antenna gain.

In order to measure the energy consumption, we analyze the average energy per packet, that is, the average energy required to deliver a packet to the BS. Formally, we need to define two functions $E_c(t)$ and $D_c(t)$ representing the amount of energy spent and the number of data packets delivered up to time slot $t$, respectively. Both are non decreasing functions, which may be also seen as reward functions of a properly designed Markov process $\mathcal{M}$.

The process $\mathcal{M}$ is characterized by a state $S_0$ and two sets of states, namely $\mathcal{S}_r$ and $\mathcal{S}_b$. The states belonging to the former set are identified by the number of packet(s) received at the relay and at the BS after the first transmission performed by the UE; the states in $\mathcal{S}_b$ are instead identified by the number of packets received at the BS after the second time slot.
The process starts in state $S_0$, and after the first time slot moves into one of the states in $\mathcal{S}_r$, depending on which packet(s) are decoded at the relay and at the BS. The cost of this transition is $TP_t$, equal to the energy consumed by the UE, being $T$ the time slot duration.
In the subsequent time slot, the process moves to one of the states in $\mathcal{S}_b$, depending on which packet(s) are decoded at the BS. This transition has a cost equal to the total power consumed by the UE and by the relay, and depends on the relaying scheme and on the success of the previous transmission.
Finally, the system goes back to $S_0$ with probability one. The reward of this last transition is equal to the overall number of packets $D$ delivered in the last two time slots, while its cost is zero.
At each step, the cost and the reward are added to $E_c(t)$ and $D_c(t)$, respectively.
Therefore, the average energy per packet $\mathcal{E}$ is defined as:
\begin{equation}
 \mathcal{E} = \mathbb{E}\left[\lim_{t\rightarrow\infty}\frac{E_c(t)}{D_c(t)}\right]
\end{equation}

According to Renewal Theory, $\mathcal{E}$ can be computed as:
\begin{equation}
 \mathcal{E} = \frac{\mathbb{E}\left[E_s\right]}{\mathbb{E}\left[D_s\right]}
\end{equation}
where $E_s$ and $D_s$ are the energy and the number of packets delivered in two subsequent time slots.

For the basic scheme, it can be easily derived that $\mathbb{E}\left[E_s\right] = 2P_tT$, whereas $\mathbb{E}\left[D_s\right] = T_{dir}$. Therefore $\mathcal{E}_{dir} = 2P_tT/T_{dir}$.

For the relayed schemes, the denominator is equal to the average throughput computed in the previous section. As to the numerator, we observe that in the \textit{Baseline relaying} scheme, the amount of energy used by the UE is again $2P_tT$. In addition, the relay energy $\bar{P}_rT$ has to be added when the relay successfully decodes the packet sent by the UE. Therefore, we get:
\begin{equation}
 \mathcal{E}_{rel} = \frac{T(2P_t + \bar{P}_r\mathbb{E}[\chi_{ur}])}{T_R}
\end{equation}
Notice that the value of $\bar{P}_r$ can be considerably lower than $P_t$ if the transmit antenna gain $\eta$ on the backhaul link is high. We also observe that the \textit{Selection} relaying scheme consumes exactly the same average energy whereas, when feedback is available, the relay transmits only if it has decoded the packet from the UE and the same packet has not been already received at the BS. Hence:
\begin{equation}
 \mathcal{E}_{rel}^s = \frac{T(2P_t + \bar{P}_r\mathbb{E}[\chi_{ur}])}{T_R^s} \quad\quad \mathcal{E}_{rel}^f = \frac{T(2P_t + \bar{P}_r\mathbb{E}[\chi_{ur}]\left(1 - \mathbb{E}[\chi_{ub1}]\right))}{T_R^f}
\end{equation}

\subsection{Impact of directional antennas}
In the previous analysis, omnidirectional antennas have been used both at the relay and at the BS. However, intuition suggests that, since relays are deployed to help users far from the cell-center, it could be beneficial to adopt a directional antenna for the access link, pointing to the cell edge.

In order to test this type of relay configuration, we use the antenna pattern introduced in (\ref{trigopattern}):
\begin{equation}
 f(\theta) = (k+1)\left(\frac{1+\cos(\theta)}{2}\right)^k
\end{equation}
where the normalization factor $k+1$ comes from the ratio $4\pi/\int_0^{2\pi}\int_0^{\pi}f(\theta)\sin(\theta)\,d\theta d\phi$ accounting for the antenna directivity. The same expressions for throughput and energy consumption can be used in this case, with the following differences:
\begin{itemize}
 \item the average SNR at the relay now depends not only on the distance $d_{ur}$ between the UE and the relay, but also on the direction $\theta_{ur}$ from which the communication is performed; therefore $\gamma_{ur} = P_tAf(\theta_{ur})/(N_0d_{ur}^{\alpha})$;
 \item the expression of $\mathscr{L}_d(s,t,x) = \mathbb{E}\left[e^{-sI_1}e^{-tI_2}\right]$, in (\ref{defLd}), is to be correspondingly corrected:
 \begin{eqnarray}
 \mathscr{L}_d(s,t,x) & = & \exp\left(-\lambda\int_0^{2\pi}\int_x^{+\infty}\left(1-\frac{1}{\left(1+tP_tAr^{-\alpha}\right)}\times\right.\right.\nonumber\\
 &\hspace{-4cm} &\hspace{-3cm} \times\left.\left.\frac{1}{\left(1+sP_tA\,f\!\left(\arcsin\left(\frac{r\sin\theta}{\sqrt{r^2+d^2-2rd\cos(\theta)}}\right)\right)\left(r^2+d^2-2rd\cos(\theta)\right)^{-\alpha/2}\right)}\right)r\,drd\theta\right)
 \label{defLdmod}
\end{eqnarray}
\end{itemize}

\section{Advantages of Superposition Coding}
\label{adv_SC}
The use of SIC is particularly effective for UEs whose channel towards the BS is not too bad. In this case, SIC allows to send additional information on this channel while the backhaul link is also being used by the relay.

A dual effect can be obtained by exploiting superposition coding at the UE. In this case, additional information can be sent over the direct link while the access link is being used. Superposition coding can be seen as an effective way to transmit two packets simultaneously, by properly sharing the transmitter resources (in this case, power) between the packets, which can then be decoded via SIC.

A UE which is close to the BS may therefore deliver two packets with a single transmission. However, UEs located farther can hardly obtain the same benefit: it may happen, in fact, that the two packets interfere too much with each other, resulting in degraded performance.
The presence of a relay close to the UE may balance this effect. If we assume that a fraction $0.5<\beta<1$ of power is reserved for packet 1, while $1-\beta$ for packet 2, a careful choice of $\beta$ can make the difference. A high value of $\beta$ may let the BS successfully decode packet 1, although packet 2 is likely to be lost due to the noise plus interference level at the BS; the relay, which is closer, has a much higher probability of decoding both packets, and could therefore forward packet 2 at the BS with full power (while the UE sends a third packet, if SIC is used at the BS)
\footnote{Several refinements could be added to the scheme, including the retransmission of a combination of packet 1 and packet 2 in case combining techniques are adopted at the BS. We leave the analysis of these variants for future work.}.

We investigate whether such an option could be advantageous, and the achievable gain.

Clearly, for a fair comparison, we also have to consider the usage of superposition coding at the UE when relaying is not adopted. In this case, we can easily derive the optimal value of $\beta$ to be chosen. In fact, the probability of decoding both packets at the BS is given by:
\begin{eqnarray}
 P_{12}^{ub} & = & P\left[\frac{\beta\gamma_{ub}h}{(1-\beta)\gamma_{ub}h+\hat{I}+1}\geq\vartheta, \frac{(1-\beta)\gamma_{ub}h}{\hat{I}+1}\geq\vartheta\right]\nonumber\\
 & = & \exp\left(-\frac{\vartheta(\hat{I}+1)}{\gamma_{ub}}\max\left(\frac{1}{\beta(\vartheta+1)-\vartheta},\frac{1}{1-\beta}\right)\right)
\end{eqnarray}
where we assume that perfect cancellation is achievable upon decoding the first packet. The maximum is to be taken between two terms depending on $\beta$: the former decreases with $\beta$, whereas the latter increases. Therefore, probability $P_{12}^{ub}$ is maximized when they have the same value, meaning that the optimal value of $\beta$ for direct transmission is:
\begin{equation}
 \beta^{opt} = \frac{\vartheta+1}{\vartheta+2}
\end{equation}

The value $\beta^{opt}$ does not depend on either the UE location or the interference level $I$, since both packets are transmitted over the same channel.

When relaying is considered, instead, the optimal choice of $\beta$ requires a different approach. The aim here is to maximize the probability that the first packet is decoded at the BS, whereas the second one (and hence both packets) is received at the relay. Since the fading on the direct link and that on the access link are independent, for fixed values of the interference levels $I_B$ and $I_R$ we have that:
\begin{eqnarray}
 P_{12}^R & = & P\left[\frac{\beta\gamma_{ub}h}{(1-\beta)\gamma_{ub}h+\hat{I}_B+1}\geq\vartheta\right]P\left[\frac{\beta\gamma_{ur}h}{(1-\beta)\gamma_{ur}h+\hat{I}_R+1}\geq\vartheta, \frac{(1-\beta)\gamma_{ur}h}{\hat{I}_R+1}\geq\vartheta\right]\nonumber\\
 & = & \exp\left(-\frac{\vartheta(\hat{I}_B+1)}{\gamma_{ub}}\frac{1}{\beta(\vartheta+1)}\right)\exp\left(-\frac{\vartheta(\hat{I}_R+1)}{\gamma_{ur}}\max\left(\frac{1}{\beta(\vartheta+1)-\vartheta},\frac{1}{1-\beta}\right)\right)
\end{eqnarray}

The first exponential term is always growing with $\beta$, while the second one is increasing only up to $\beta = \beta^{opt}$, meaning that the optimal value of $\beta$ for the relaying scheme lies between $\beta^{opt}$ and 1. In this interval, $\beta(\vartheta+1)-\vartheta\geq1-\beta$, leading to:
\begin{equation}
  P_{12}^R = \exp\left(-\frac{\vartheta(\hat{I}_B+1)}{\gamma_{ub}}\frac{1}{\beta(\vartheta+1)-\vartheta}-\frac{\vartheta(\hat{I}_R+1)}{\gamma_{ur}}\frac{1}{1-\beta}\right)
\end{equation}
which depends also on the interference levels $I_B$ and $I_R$. Unless these values are known at the UE, the expectation has to be taken over both $I_B$ and $I_R$:
\begin{equation}
 P_{12}^R = e^{-\frac{\vartheta}{\gamma_{ub}}\frac{1}{\beta(\vartheta+1)-\vartheta}} e^{-\frac{\vartheta}{\gamma_{ur}}\frac{1}{1-\beta}}\mathscr{L}_d\left(\frac{\vartheta}{N_0\gamma_{ur}(1-\beta)},\,\frac{\vartheta}{N_0\gamma_{ur}(\beta(\vartheta+1)-\vartheta)},d_{ub}\right)
 \label{valP12}
\end{equation}

The optimal value $\beta_R^{opt}$ is therefore difficult to derive, since the closed form expression of $P_{12}^R$ is not available. As a first approximation, it is possible to rely on the SNRs, rather than the SINRs, thus ignoring the interference. In this case, $P_{12}^R$ reduces to the first two exponential terms in (\ref{valP12}), and the optimal $\beta$ can be easily found as:
\begin{equation}
 \beta_R^{opt} = \arg\min_{\beta}\left(\frac{1}{\gamma_{ub}[\beta(\vartheta+1)-\vartheta]} + \frac{1}{\gamma_{ur}(1-\beta)}\right)
\end{equation}
Through simple calculations, and having in mind the lower bound on $\beta_R^{opt}$, we finally derive:
\begin{equation}
 \beta_R^{opt} = \max\left(1-\frac{1-\sqrt{\frac{1}{\vartheta+1}\left(\frac{d_{ub}}{d_{ur}}\right)^{\alpha}}}{\vartheta+1-\left(\frac{d_{ub}}{d_{ur}}\right)^{\alpha}},\,\, \frac{\vartheta+1}{\vartheta+2}\right)
\label{val_betaR_opt}
\end{equation}
where the maximum is to be taken, since the first term is bounded between $\vartheta/(\vartheta+1)$ and 1.

We stress two important observations. First, if the values of $I_B$ and $I_R$ at each time slot were known, an upper bound on the system performance could be obtained, since $\beta_R^{opt}$ can be always selected without approximation through equation (\ref{val_betaR_opt}), where $\left(d_{ub}/d_{ur}\right)^{\alpha}$ is replaced with $\left(d_{ub}/d_{ur}\right)^{\alpha}\frac{\hat{I}_B+1}{\hat{I}_R+1}$. Secondly, $\beta^{opt}$ and $\beta_R^{opt}$ are likely to be different: this means that opportunistic schemes are less beneficial. In other words, without SC it is possible for the UE to transmit the packet and fully exploit the spatial diversity between the channels towards the BS and the relay (with the \textit{Selection} and \textit{Feedback} schemes described above).
When SC is employed, instead, the UE has to choose in advance the value of $\beta$ which maximizes the success probability on either the direct or the access link.

The throughput expressions when SC is adopted can be found in a way analogous to that applied to the simpler case. However, we need to introduce some new binary random variables, taking into account the transmissions of two superimposed packets ($x$ and $y$) on the same channel:
\begin{itemize}
 \item $\chi_{ub}^x=1$ if only one of the two superimposed packets on the direct link is decoded by the BS;
 \item $\chi_{ub}^y=1$ if both superimposed packets on the direct link are successfully decoded at the BS;
 \item $\chi_{ur}^y=1$ if both superimposed packets on the access link are successfully decoded at the relay;
\end{itemize}

It follows that we have:
\begin{eqnarray}
 T_{dir}^{SC}(d_{ub},\theta_u) & = & 4\mathbb{E}\left[\chi_{ub}^y\right]\label{TdirSC}\\
 T_R^{SC}(d_{ub},\theta_u) & = & \mathbb{E}\left[\chi_{ub}^x\right] + \mathbb{E}\left[\chi_{ub}\right] + \mathbb{E}\left[\chi_{ur}^y\chi_{rb}\right] - \mathbb{E}\left[\chi_{ur}^y\chi_{ub}\right] + \mathbb{E}\left[\chi_{ur}^y\chi_{ub}^i\right]\label{TrelSC}\\
 T_R^{sSC}(d_{ub},\theta_u) & = & T_R^{SC}(d_{ub},\theta_u) + \mathbb{E}\left[\chi_{ub}^y\right] - \mathbb{E}\left[\chi_{ub}^y\chi_{ur}^y\chi_{rb}\right]\label{TrelsSC}\\
 T_R^{fSC}(d_{ub},\theta_u) & = & T_R^{sSC}(d_{ub},\theta_u) - \mathbb{E}\left[\chi_{ub}^y\chi_{ur}^y\chi_{ub}^i\right] + \mathbb{E}\left[\chi_{ub}^y\chi_{ur}^y\chi_{ub}\right]\label{TrelfSC}
\end{eqnarray}

\section{Results}
\label{results}
In this section, we present the performance of the proposed schemes in terms of throughput and energy consumption. In the analyzed scenarions we put $\lambda=4.6\times10^{-6}\,m^{-2}$, $A = 10^{-3}$, $\alpha = 3.7$, $N_0=-103\,dBm$.
Note that the resulting BS density is equal to that of a hexagonal grid with inter-site distance (ISD) of $500\,m$. However, since the BSs (and the UEs) are distributed according to a PPP, it is possible that a UE is deployed much farther than $250\,m$ from its BS. Unless specified, the UE transmission power $P_t$ is equal to $23\,dBm$.

\begin{figure}
    \centering
    \includegraphics[width=\figw]{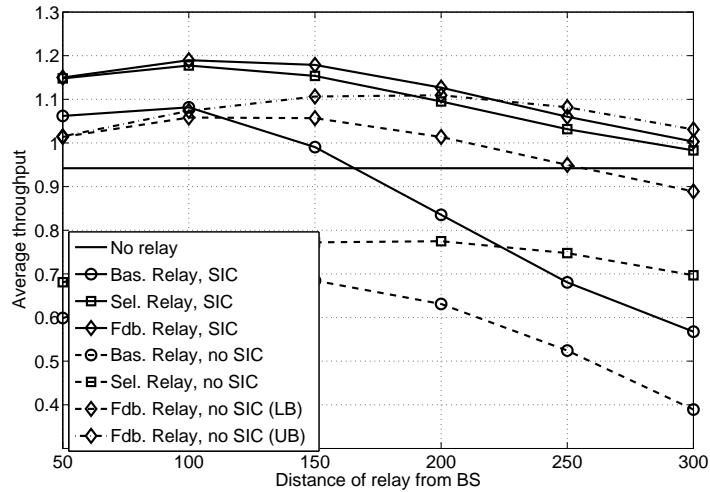}
     \caption{\small Average throughput as a function of the distance of the relays from the BS, when three relays are deployed and $P_r = 2P_t$, with and without the employment of a SIC receiver at the BS.}
     \vspace{-1cm}
  \label{fig:Ave_thro_vardr_2Pr_vsSIC}
\end{figure}

We plot the average throughput as a function of $d_{rb}$ in Figure \ref{fig:Ave_thro_vardr_2Pr_vsSIC}. We observe that for all the relaying schemes there is an optimal relay location. A relay too close to the BS grants little benefit to UEs located far from the cell center, whereas, on the contrary, if $d_{rb}$ is too large the backhaul link becomes too weak to improve the throughput of cell-edge users.
We also notice that if the UE is allocated to the relay, regardless of its position, the performance soon degrades as the relay is located farther from the BS. This is due to the fact that UEs close to the center of the cell often do not need relaying, which on the contrary appears to be detrimental. A significant advantage is observed when a selection between the direct and the relayed transmission is made.
In this case, only UEs which actually need relaying are helped, bringing a throughput improvement of almost 20\%. Finally, if an error-free feedback is available at the end of the first slot, some additional performance gain is available. This is due to the fact that unnecessary relay transmissions are performed, hence reducing the interference at the BS. 

An interesting question is whether the use of SIC actually has a beneficial or a detrimental effect in terms of throughput. In fact, one may argue that the second transmission from the UE may turn out to be a strong interference to the relay signal rather than a way to deliver a second packet. This is why we also plot in Figure \ref{fig:Ave_thro_vardr_2Pr_vsSIC} the throughput, as a function of the relay location, when SIC is not employed.
In this case, for both the \textit{Baseline} relaying scheme and the \textit{Selection} relaying scheme, the UE is silent during the second time slot, allowing the relay to forward the packet sent in the previous time slot. If feedback from the BS is available, in case of reception of the first packet at the BS, the relay does not forward, and instead the UE transmits also in the second time slot.
We notice that this choice also changes the interference level. Assuming that the network is coordinated, all the relays transmit in the same time slots. Therefore, if the \textit{Baseline} relaying scheme or the \textit{Selection} relaying scheme is adopted in the network, when the relay forwards only the other relays are transmitting, which results in a negligible interference from the other cells, according to our assumptions.
This is not true when feedback is used, or when, in general, some UEs are relayed and the other ones transmit directly. In this case, the computation of the interference during the relay transmission is not straightforward.
In general, we use a lower and an upper bound: the former is obtained by considering that all the UEs of the other cells are not relayed, and therefore they all transmit in all time slots (just as when SIC is adopted), while the latter is obtained by assuming that all the UEs are relayed, and therefore no interference is assumed when the relay forwards.

We observe that the usage of SIC always outperforms the schemes obtained without it, despite the lower interference during the relay transmission. In particular, the usage of half duplex relays is detrimental if neither feedback nor SIC are exploited, due also to the fact that UEs not too far from the BS suffer from the reduced transmission time available.
As regards the \textit{Feedback} scheme without SIC, it allows an improvement with respect to the basic scheme, but the benefit is less than that achievable with SIC. The upper and lower bounds show a larger gap when $d_{ur}$ increases, since the different level of interference considered on the backhaul link has a stronger effect when this link is weaker.

\begin{figure}
    \centering
    \includegraphics[width=\figw]{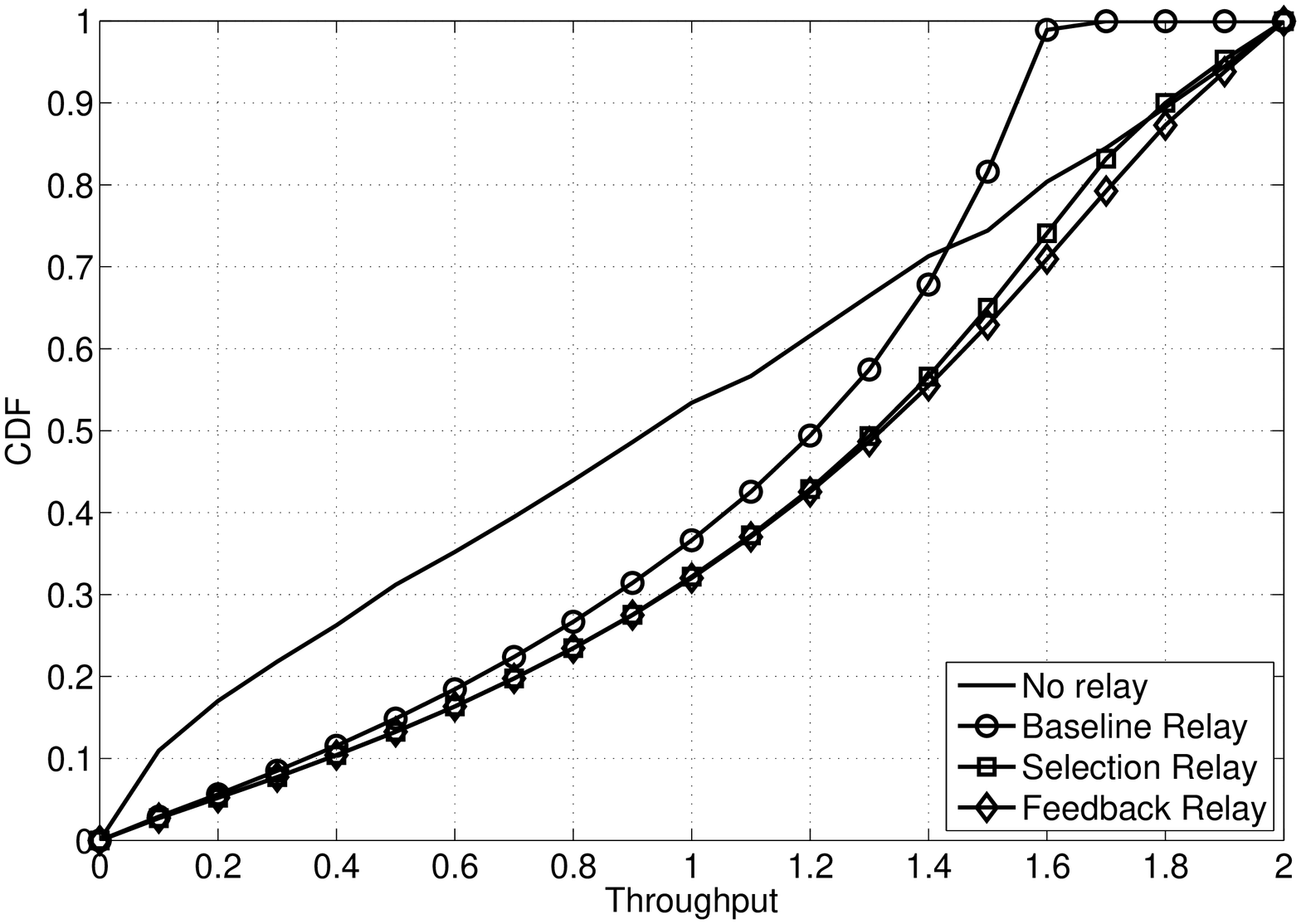}
    \vspace{-0.5cm}
     \caption{\small Throughput CDF, when the distance $d_{rb}$ between the relay and the BS is equal to $150\,m$, for the 4 different schemes. Here the effective transmit power at the relay is $2P_t$.}
  \label{fig:CDFThro_SIC_150m}
\end{figure}

We also report in Figure \ref{fig:CDFThro_SIC_150m} the CDF of the throughput, for the four schemes. We observe, as expected, that the \textit{Baseline} relaying scheme performs better for cell edge UEs, which can highly increase their performance, but in turn is detrimental for UEs located close to the BS, which are forced to have their packets relayed. The usage of more refined schemes, such as the \textit{Selection} and the \textit{Feedback} scheme, can avoid this drawback.

\begin{figure}
    \centering
    \includegraphics[width=\figw]{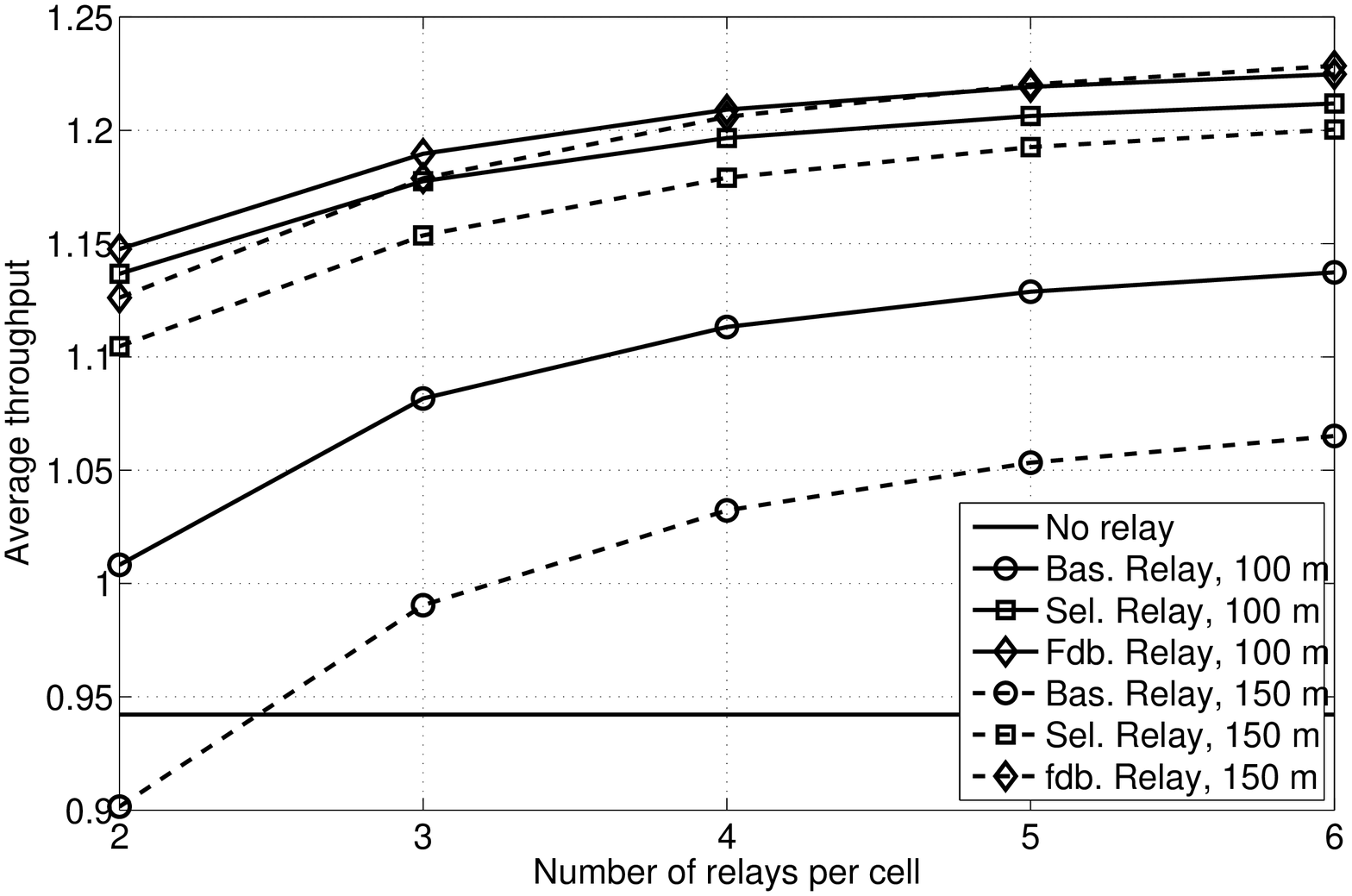}
    \vspace{-0.5cm}
     \caption{\small Average throughput as a function of the number of relays per cell, for two different relay distances from the BS (100 and 150 m). Here the effective transmit power at the relay is $2P_t$.}
     \vspace{-1cm}
  \label{fig:Ave_thro_varrel}
\end{figure}

Secondly, we investigate the impact of the number of relays per cell, from 2 up to 6, in Figure \ref{fig:Ave_thro_varrel}. The computation can be done by properly setting the value $k_r$ in (\ref{avethro}).
We clearly observe that the performance improves with the number of relays, as expected, at the cost of an increased deployment cost.
However, the improvement is more remarkable for few relays (whereas no substantial throughput increase can be obtained with $k_r>5$) and is shown by all the adopted schemes. In the figure, we focus on two relay distances, namely 100 and 150 meters, which were found to be the best in Figure \ref{fig:Ave_thro_vardr_2Pr_vsSIC}.
We notice also that, when the number of relays is large enough, and feedback is available, the best position for the relays tends to shift farther from the cell center. This is due to the fact that when few relays are available, placing them away from the BS prevents them from fully covering the cell area, which is not true when more relays can be deployed.

\begin{figure}
    \centering
    \includegraphics[width=\figw]{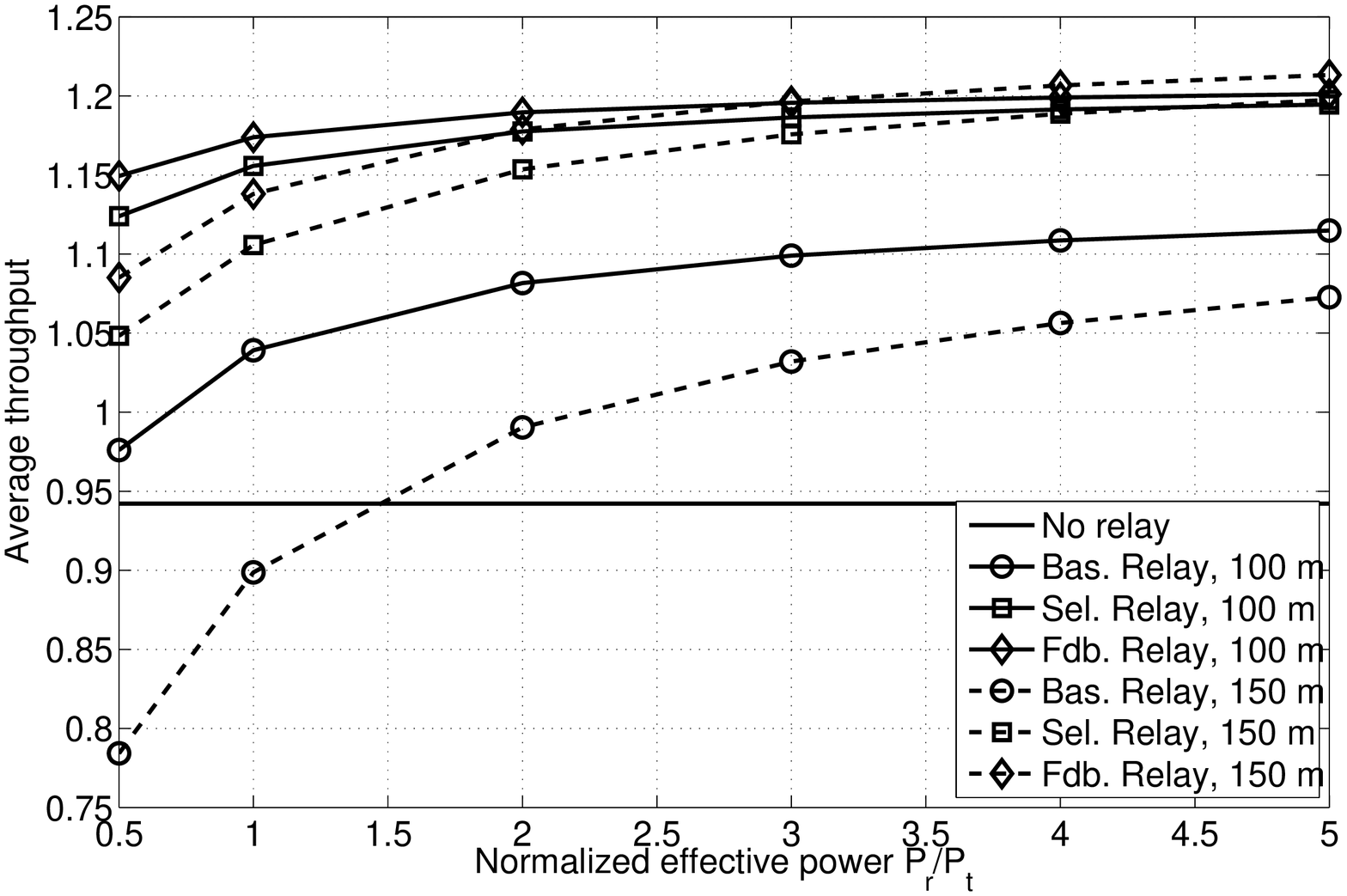}
     \caption{\small Average throughput as a function of the relay transmission power, for two different relay distances from the BS (100 and 150 m). Here 3 relays are deployed per cell.}
     \vspace{-1cm}
  \label{fig:Ave_thro_varPr}
\end{figure}

We study in Figure \ref{fig:Ave_thro_varPr} the effect of the relay transmission power. More precisely, we plot the average throughput achieved with different effective relay power $P_r$, normalized to the UE transmission power $P_t$.
As intuition suggests, the performance improve with the relay power, due to our assumption that the relays cause negligible interference to other BS. However, the performance curves tend to become flat for all schemes as $P_r$ increases. In fact, a power of $2P_t$ is enough to guarantee that the backhaul link is reliable, and any further power increase adds little benefit.
As before, we observe that as the relay transmission power becomes higher, the best relay position shifts away from the cell center, since longer backhaul links are still strong. This is true when feedback or channel selection is available, since otherwise the throughput loss of the UEs close to the cell center prevents further improvements.

\begin{figure}
    \centering
    \includegraphics[width=\figw]{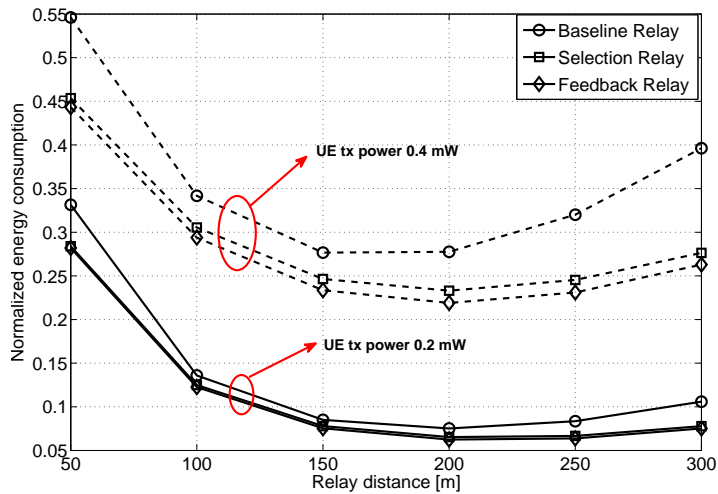}
     \caption{\small Normalized energy consumption as a function of the distance of the relays from the BS, when three relays are deployed, $\eta=10$, and $P_r = 2P_t$.}
     \vspace{-1cm}
  \label{fig:Enereffi_SIC}
\end{figure}

We study the energy consumption in Figure \ref{fig:Enereffi_SIC}, where we plot the normalized energy consumption (namely, the ratio between the energy per packet spent with relaying and the one spent with the basic scheme) as a function of the distance of the relays from the BS. The effective relay transmit power is fixed to $P_r = 2P_t$, whereas the relay antenna gain on the backhaul link is $\eta = 10 dB$.
We stress that while $P_r$ has an impact on the throughput, $\eta$ determines the actual energy consumed at the relay. We observe that for all the relay schemes the energy consumption per packet is much lower. This is due to the fact that the coverage holes are covered by means of the relay, thus highly reducing the number of retransmissions needed for cell edge users. This is confirmed by the fact that, when the transmit power $P_t$ is doubled, the gain achieved via relaying is lower.
Notice however that further increasing $P_t$ is not likely to improve the performance very much, since the system quickly becomes interference limited.
As regards the effect of the relay location, we notice that the best position is somewhere half way between the BS and the cell edge. In this case, not only are the cell edge users well served, but also the backhaul link is still strong enough to efficiently deliver the relayed packets.
The schemes with selection and with feedback can further decrease the energy expense. However, the impact of feedback is quite negligible. This is due to the fact that the highest improvement is for cell edge users, which have little benefit from the use of feedback. In fact, when $P_t$ is doubled, also the gap between the performance of \textit{Selection} relaying scheme and \textit{Feedback} relaying scheme is more remarkable.

\begin{figure}
    \centering
    \includegraphics[width=\figw]{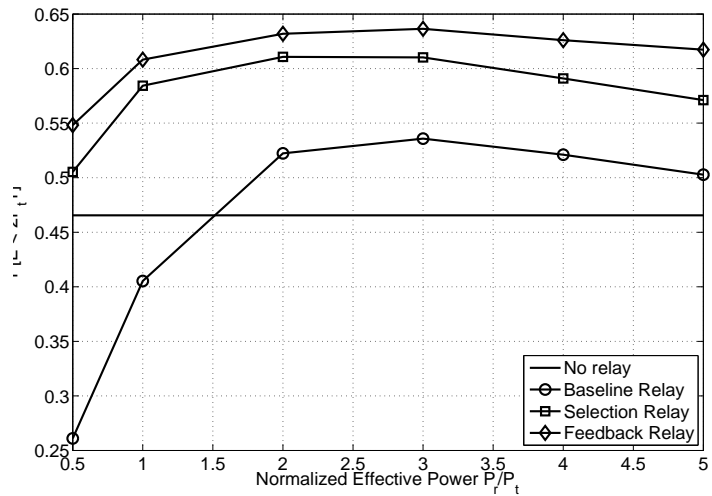}
     \caption{\small Probability of having an energy consumption lower than $2P_tT$ per packet, when three relays are deployed at 150 m from the BS and $\eta=10$, as a function of $P_r$.}
     \vspace{-1cm}
  \label{fig:Prob_lowener2Pt_150m_varPr}
\end{figure}

The augmented relay transmission power $P_r$, provided that the antenna gain $\eta$ is fixed, implies that the higher throughput comes at the cost of a higher amount of energy spent. The consequence is that a balance between throughput and energy has to be found.
The average energy consumption is not a useful metric, since, especially when the relay is not too far from the BS, the main contribution is brought by the UEs located at the cell edge, whose performance is almost unaffected by an increased relay transmission power.
It may be useful instead to observe the behavior of the UEs which are in the cell center, and which show, therefore, a low energy consumption. We focus on the scenario with 3 relays per cell, located at 150 m. After numerically computing the energy consumption distribution, we plot in Figure \ref{fig:Prob_lowener2Pt_150m_varPr} the probability that the energy consumption per packet is lower than a threshold $E_{\ell} = 2P_tT$ (meaning that a single retransmission is enough to deliver the packet without the help of a relay) as a function of the relay effective transmission power.
It can be observed that the probability of a low energy consumption increases with $P_r$, due to the fact that the backhaul link becomes stronger. However, when $P_r$ grows too much, the cost in terms of energy becomes higher than the benefit in terms of energy saving, thus leading to a lower performance.
The usage of the relays in any case improves the number of UEs with low energy consumption, provided that the more refined schemes are used. \textit{Baseline} relaying scheme performs worse than direct transmission since a UE close to the BS is forced to have half of its packets relayed, thus lowering its throughput.
The conclusion is that, in order to minimize the energy per packet, the effective transmit power $P_r$ should be chosen dynamically, based on the UE location.

\begin{figure}
    \centering
      \psfrag{aaa}{\small$\pi/2$}
      \psfrag{bbb}{\small$\pi/3$}
      \psfrag{ccc}{\small$\pi/4$}
      \psfrag{ddd}{\small$\pi/6$}
     \includegraphics[width=\figw]{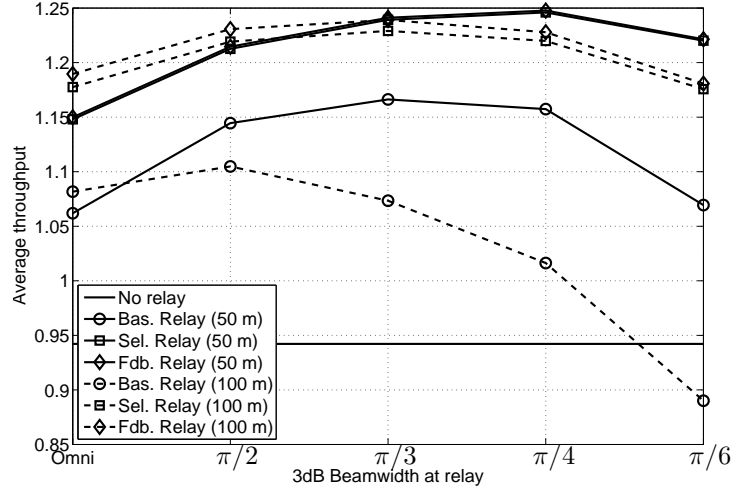}
     \caption{\small Average throughput as a function of the 3dB beamwidth of the relay antenna, for two different relay distances from the BS (50 and 100 m). Here 3 relays are deployed per cell, with effective transmission power equal to $2P_t$.}
  \label{fig:Ave_thro_varang}
\end{figure}

When directional antennas are used at the relays, the performance can be further improved, as observed in Figure \ref{fig:Ave_thro_varang}. Here, we plot the average throughput for different values of the antenna beamwidth at the relay, and for two locations of the relay, at 50 and at 100 meters away from the BS. The throughput is increased, provided that the beamwidth is not too small, due to the fact that the relays are able to better cover the fraction of the cell they are dedicated to.
In addition, we notice that the best position of the relay tends to shift towards the BS as the beamwidth is decreased. This is due to the fact that the directional antenna makes it more difficult for UEs located in the area not covered by the antenna to efficiently communicate with the relay. Placing the relay closer to the center of the cell actually reduces the area where this effect is more pronounced.

\begin{figure}
    \centering
     \includegraphics[width=\figw]{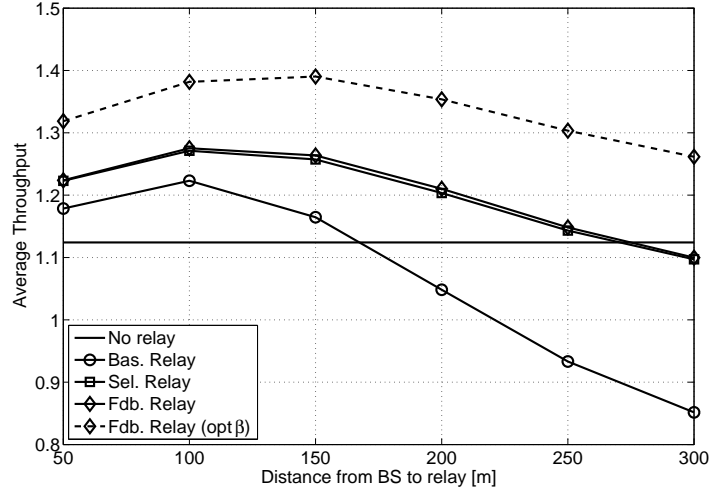}
     \vspace{-0.5cm}
     \caption{\small Average throughput as a function of the distance between the relay and the BS. Here 3 relays are deployed per cell, with effective transmission power equal to $2P_t$.}
  \label{fig:Ave_throSC_vardr_2Pr}
\end{figure}

\begin{figure}
    \centering
    \includegraphics[width=\figw]{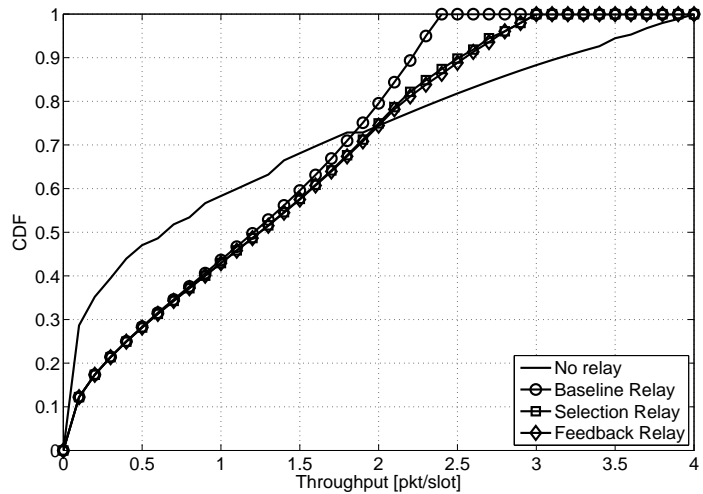}
      \vspace{-0.5cm}
     \caption{\small Throughput CDF with Superposition Coding, when the distance $d_{rb}$ between the relay and the BS is equal to $150\,m$, for the 4 different schemes. Here the effective transmit power at the relay is $2P_t$.}
  \label{fig:CDFThroSC_SIC_150m}
  \vspace{-1cm}
\end{figure}

Finally, we consider the improvement achievable via Superposition Coding. We focus on the basic scenario, with $k_r=3$ relays and $P_r = 2P_t$. First, we plot in Figure \ref{fig:Ave_throSC_vardr_2Pr} the average throughput as a function of the distance $d_{rb}$.
Clearly, even the basic scheme shows an improved performance, since the UE can deliver up to two packets to the BS per time slot. Nevertheless, relaying is still able to grant better performance, even with the baseline relaying scheme, provided that the relays are not deployed too far from the BS.
The relative gain, however, is lower. The main reason is the fact that the opportunistic receiver selection used in both the \textit{Selection} scheme and the \textit{Feedback} scheme is less effective. In fact, the choice of the optimal $\beta$ at the UE would be different for the direct link and the access link. The relaying scheme optimizes $\beta$ for the relayed link, which however is suboptimal if the UE is located in the cell center.
This can be clearly seen in Figure \ref{fig:CDFThroSC_SIC_150m}, where the throughput CDF is depicted for the case $d_{rb} = 150\,m$.
All the relaying schemes are beneficial for UEs at the cell edge, whose throughput is enhanced. In this case, optimizing $\beta$ for the relayed channel is the best choice. On the contrary, direct transmission outperforms relaying for the UEs located in the cell center, since it optimizes $\beta$ without considering the presence of the relay, which is actually useless in this case.

It follows that the best choice is to optimize $\beta$ by selecting either $\beta^{opt}$ or $\beta_R^{opt}$, depending on which one offers the best performance for the specific UE location. This is the curve marked as \textit{Fdb. Relay (opt. $\beta$)} in Figure \ref{fig:Ave_throSC_vardr_2Pr}, where we average the function $\max(T(d_{ub},\theta_u), T_R^f(d_{ub},\theta_u))$ for every possible location $(d_{ub},\theta_u)$ of the UE. As can be observed (and even if $\beta_R^{opt}$ is computed via (\ref{val_betaR_opt}), hence ignoring interference), the resulting performance is by far the best one achievable through relaying.

\section{Conclusions}
\label{concl}
In this paper, we have analyzed the performance gain, in terms of overall throughput and energy consumption, achievable when relays are used for uplink transmission. Inter-cell interference has been taken into account, and several factors have been considered, such as the number of relays per cell, their available transmit power, their location and antenna beam pattern. Numerical simulations showed that relays can be useful not only for cell-edge users, but also for mobiles located closer to the Base Stations.
The usage of more refined decoding schemes, based on Successive Interference Cancellation and on Superposition Coding, can in fact allow the transmission of parallel communications over the direct and the relayed path from the mobile to the BS. A proper setting of the parameters can therefore grant an average throughput increase of almost 40\%.

\appendix

The computation of $\mathbb{E}\left[\chi_{ub1}\chi_{ur}\chi_{rb}\right]$ follows the main line adopted in (\ref{E_urrb}), with the only difference that three links are to be considered now. As before, we observe that packet decoding on the backhaul link must consider the fact that there is a concurrent transmission from the UE to the BS, and hence SIC is necessary; it follows that two cases are possible, depending on the decoding order at the BS.

Therefore, we compute:
\begin{eqnarray}
 \mathbb{E}\left[\chi_{ub1}\chi_{ur}\chi_{rb}\right] & = &  \mathbb{E}_{I_{B1},I_R,I_{B2}}\left[P\left[\frac{\Gamma_{ub1}}{\hat{I}_{B1}+1}\geq\vartheta,\frac{\Gamma_{ur}}{\hat{I}_R+1}\geq\vartheta,\frac{\Gamma_{rb}}{\Gamma_{ub}+\hat{I}_{B2}+1}\geq\vartheta\right]+\right.\nonumber\\
& & + \left.P\left[\frac{\Gamma_{ub1}}{\hat{I}_{B1}+1}\geq\vartheta,\frac{\Gamma_{ur}}{\hat{I}_R+1}\geq\vartheta,\frac{\Gamma_{ub2}}{\Gamma_{rb}+\hat{I}_{B2}+1}\geq\vartheta, \frac{\Gamma_{rb}}{\hat{I}_{B2}+1}\geq\vartheta\right]\right] \nonumber\\
 & = & e^{-\vartheta\left(\frac{1}{\gamma_{ub}}+\frac{1}{\gamma_{ur}}+\frac{1}{\gamma_{rb}}\right)}\frac{1}{1+\frac{\gamma_{ub}}{\gamma_{rb}}\vartheta}\,\mathbb{E}_{I_{B1},I_R,I_{B2}}\left[e^{-\frac{\vartheta}{\gamma_{ub}}\hat{I}_{B1}}e^{-\frac{\vartheta}{\gamma_{ur}}\hat{I}_R}e^{-\frac{\vartheta}{\gamma_{ub}}\hat{I}_{B2}}\right] + \nonumber\\
 & & +\, e^{-\vartheta\left(\frac{1}{\gamma_{ur}}+\frac{1}{\gamma_{rb}}+\frac{2}{\gamma_{ub}}\right)}\frac{e^{-\frac{\vartheta^2}{\gamma_{ub}}}}{1+\frac{\gamma_{rb}}{\gamma_{ub}}\vartheta}\,\mathbb{E}_{I_{B1},I_R,I_{B2}}\left[e^{-\frac{\vartheta}{\gamma_{ub}}\hat{I}_{B1}}e^{-\frac{\vartheta}{\gamma_{ur}}\hat{I}_R}e^{-\frac{\vartheta(\vartheta+1)}{\gamma_{ub}}\hat{I}_{B2}}e^{-\frac{\vartheta}{\gamma_{rb}}\hat{I}_{B2}}\right] \nonumber \\
 & = & \frac{e^{-\vartheta\left(\frac{1}{\gamma_{ub}}+\frac{1}{\gamma_{ur}}+\frac{1}{\gamma_{rb}}\right)}}{1+\frac{\gamma_{ub}}{\gamma_{rb}}\vartheta}\mathscr{L}_{d_{rb},0}\left(\frac{\vartheta}{N_0\gamma_{ur}},\frac{\vartheta}{N_0\gamma_{ub}},\frac{\vartheta}{N_0\gamma_{rb}}, d_{ub}\right) +\nonumber\\
 & & +\, \frac{e^{-\vartheta\left(\frac{1}{\gamma_{ur}}+\frac{1}{\gamma_{rb}}+\frac{(\vartheta+2)}{\gamma_{ub}}\right)}}{1+\frac{\gamma_{rb}}{\gamma_{ub}}\vartheta}\mathscr{L}_{d_{rb},0}\left(\frac{\vartheta}{N_0\gamma_{ur}},\frac{\vartheta}{N_0\gamma_{ub}},\frac{\vartheta}{N_0}\left(\frac{\vartheta+1}{\gamma_{ub}}+\frac{1}{\gamma_{rb}}\right),d_{ub}\right)
 \label{E_ub1urrb}
\end{eqnarray}
where we indicated with $\hat{I}_{B1}$ and $\hat{I}_{B2}$ the normalized interference levels at the BS during the first and the second time slot. We also introduced $\mathscr{L}_{d_{rb},0}\left(s,t,u,x\right) = \mathbb{E}\left[e^{-sI_1}e^{-tI_2}e^{-uI_3}\right]$, which is the correlation between the interference level $I_1$ measured at point $(d_{rb},0)$ in time slot $t$, the interference level $I_2$ measured at the origin in the same time slot and the interference level $I_3$ measured at the origin in time slot $t+1$.
Extending the expression in (\ref{defLd}), we get:
\begin{eqnarray}
 \mathscr{L}_d(s,t,u,x) & = & \exp\left(-\lambda\int_0^{2\pi}\int_x^{+\infty}\left(1-\frac{1}{\left(1+sP_tA\left(r^2+d^2-2rd\cos(\theta)\right)^{-\alpha/2}\right)}\times\right.\right. \nonumber\\
 & & \times \left.\left. \frac{1}{\left(1+tP_tAr^{-\alpha}\right)\left(1+uP_tAr^{-\alpha}\right)}\right)r\,drd\theta\right)
 \label{defLd3}
\end{eqnarray}

The expression of $\mathbb{E}\left[\chi_{ub1}\chi_{ur}\chi_{ub2}^i\right]$ can be obtained analogously, whereas $\mathbb{E}\left[\chi_{ub1}\chi_{ur}\chi_{ub2}\right]$ is even simpler, since SIC is not employed, resulting in:
\begin{eqnarray}
 \mathbb{E}\left[\chi_{ub1}\chi_{ur}\chi_{ub2}\right] & = & e^{-\vartheta\left(\frac{1}{\gamma_{ur}}+\frac{2}{\gamma_{ub}}\right)}\mathscr{L}_{d_{rb}}\left(\frac{\vartheta}{N_0\gamma_{ur}},\frac{\vartheta}{N_0\gamma_{ub}},\frac{\vartheta}{N_0\gamma_{ub}},d_{ub}\right)
\end{eqnarray}

\bibliographystyle{IEEEtran}
\bibliography{IEEEabrv,stabib.bib}
\end{document}